\documentclass[a4paper,10pt,twoside]{cpc-hepnp}
\usepackage{graphicx}
\usepackage{epstopdf}
\usepackage{multicol}
\usepackage{booktabs}
\usepackage{amssymb,bm,mathrsfs,bbm,amscd}
\usepackage[tbtags]{amsmath}
\usepackage{lastpage}
\usepackage{footnote}
\usepackage{CJK}
\usepackage[switch,pagewise]{lineno}
\begin{document}
\begin{CJK*}{GBK}{song}


\fancyhead[c]{\small Chinese Physics C~~~Vol. 37, No. 1 (2013)
010201} \fancyfoot[C]{\small 010201-\thepage}

\footnotetext[0]{Received 14 March 2009}

\title{Prospects for a Multi-TeV Gamma-ray Sky Survey with the LHAASO Water Cherenkov Detector Array\thanks{Supported by National Natural Science Foundation of China (11761141001, 11635011, 11873005) }}
\author{
V. Alekseenko $^ {24}$\email{yyguo@ihep.ac.cn},
Q. An $^ {4}$\email{changxc@ihep.ac.cn},
Axikegu $^ {17}$\email{huhb@ihep.ac.cn},
B. Bai $^ {18}$\email{yaozg@ihep.ac.cn},
L.X.  Bai $^ {19}$,
Y.W. Bao $^ {13}$,
D. Bastieri $^ {8}$,
X.J. Bi $^ {1,2,3}$,\\
Zhe Cao $^ {4}$,
Zhen Cao $^ {1,2,3}$
% ...
\\(LHAASO Collaboration)
}

\maketitle
\address{%
$^1$ Institute of High Energy Physics, Chinese Academy of Sciences, Beijing 100049, China\\
$^2$ University of Chinese Academy of Sciences,  Beijing 100049, China\\
$^3$TIANFU Cosmic Ray Research Center, Chengdu, Shichuan,  China\\
$^4$University of Science and Technology of China, 230026 Hefei, Anhui, China\\
$^5$Tsinghua University, 100084 Beijing, China\\
$^6$National Astronomical Observatories, Chinese Academy of Sciences, 100101 Beijing, China\\
$^7$National Space Science Center, Chinese Academy of Sciences, 100190 Beijing, China\\
$^8$Center for Astrophysics, Guangzhou University, 510006 Guangzhou, Guangdong, China\\
$^9$Sun Yat-sen University, 519000 Zhuhai, Guangdong, China\\
$^{10}$Shool of Physics and Technology,Guangxi University, 530004  Nanning, Guangxi, China\\
$^{11}$Hebei Normal University, 050024 Shijiazhuang, Hebei, China\\
$^{12}$School of Physics and Engineering, Zhengzhou University, 450001 Zhengzhou, Henan, China\\
$^{13}$Nanjing University, 210023 Nanjing, Jiangsu, China\\
$^{14}$Key Laboratory of Dark Matter and Space Astronomy, Purple Mountain Observatory, Chinese Academy of Sciences, 210034 Nanjing, Jiangsu, China\\
$^{15}$Institute of Frontier and Interdisciplinary Science, Shandong University, 266237 Qingdao, Shandong, China\\
$^{16}$Shanghai Astronomical Observatory, Chinese Academy of Sciences, 200030 Shanghai, China\\
$^{17}$School of Physical Science and Technology, Southwest Jiaotong University, 610031 Chengdu, Sichuan, China\\
$^{18}$Chengdu University of Technology, 610059 Chengdu, Sichuan, China\\
$^{19}$Sichuan University, 610065 Chengdu, Sichuan, China\\
$^{20}$Nankai University, 300071 Tianjin, China\\
$^{21}$Key Laboratory of Cosmic Rays (Tibet University), Ministry of Education, 850000 Lhasa, Tibet, China\\
$^{22}$Yunnan University, 650091 Kunming,Yunnan,China\\
$^{23}$Yunnan Astronomical Observatories, Chinese Academy of Sciences, 650216 Kunming, Yunnan, China\\
$^{24}$Institute for Nuclear Research, Russian Academy of Sciences, Moscow, Russia\\
$^{25}$D$\acute{e}$partement de Physique Nucl$\acute{e}$aire et Corpusculaire, Facult$\acute{e}$ de Sciences, Universit$\acute{e}$ de Gen$\acute{e}$ve, Geneva, Switzerland\\

}

\begin{abstract}
The Water Cherenkov Detector Array (WCDA) is a major component of the Large High Altitude Air Shower Array Observatory (LHAASO), a new generation cosmic-ray experiment with unprecedented sensitivity, currently under construction. The WCDA is aimed at the study of TeV $\gamma$-rays. In order to evaluate the prospects of searching for TeV $\gamma$-ray sources with the WCDA, we present in this paper a projection for the one-year sensitivity of the WCDA to TeV $\gamma$-ray sources from TeVCat\footnote{http://tevcat.uchicago.edu} using an all-sky approach. Out of 128 TeVCat sources observable to the WCDA up to a zenith angle of $45^\circ$, we estimate that 42 would be detectable for one year of observations at a median energy of 1 TeV. Most of them are Galactic sources, and the extragalactic sources are Active Galactic Nuclei (AGN).
\end{abstract}


\begin{pacs}
1--3 PACS(Physics and Astronomy Classification Scheme, http://www.aip.org/pacs/pacs.html/)
\end{pacs}

\footnotetext[0]{\hspace*{-3mm}\raisebox{0.3ex}{$\scriptstyle\copyright$}2013
Chinese Physical Society and the Institute of High Energy Physics
of the Chinese Academy of Sciences and the Institute
of Modern Physics of the Chinese Academy of Sciences and IOP Publishing Ltd}%

\begin{multicols}{2}

\section{Introduction}
Very high energy (VHE, $ >$100 GeV) $\gamma$-rays open a crucial window to explore the non-thermal phenomena in the Universe in their most extreme environments. Their detailed observation allows us to comprehend puzzles in modern astrophysics and cosmology, particularly for the origin of Galactic and extragalactic cosmic rays, the acceleration and radiation process in violent environments like supernova remnant (SNR) shocks, active galactic nuclei outflows or pulsar winds. Besides, it may contribute to cosmological issues by constraining the annihilation cross section of dark matter like WIMPs and searching for Lorentz invariance violation.\par
The astrophysical $\gamma$-ray sky is usually decomposed into individually-detected sources and diffuse $\gamma$-ray emission. The former, including point sources and extended sources, contains many different types: Galactic sources like SNRs, pulsar wind nebulae (PWN), binaries, etc., and extragalactic sources like AGN. Galactic cosmic rays (GCRs) are accelerated by shock waves generated in SNR \cite{Morlino2013}; electrons gain energy effectively at the termination shock of PWN where the pulsar wind is terminated by the surrounding gas, emitting TeV $\gamma$-rays via inverse Compton scattering \cite{Saha2015}\cite{DeJager1992}. Beyond our Galaxy, almost all known TeV $\gamma$-ray sources are AGN and their $\gamma$-ray emissions are thought to originate from one or multiple regions of particle acceleration in the jets. For the diffuse $\gamma$-ray emission, it is mainly attributed to the interactions of CR electrons and nuclei with interstellar gas \cite{Ackermann2012} and photon fields in the Galactic plane, providing key insight into the character of propagation of CRs in the Galaxy.\par
Various techniques have been developed to detect very high energy (VHE) $\gamma$-rays. The \textit{Fermi} Large Area Telescope (\textit{Fermi}-LAT), representing the space-borne observatories, has found thousands of  $\gamma$-ray sources in the GeV band. However, their limited effective area and low $\gamma$-ray flux at higher energies make space detectors insensitive compared with ground-based observatories in the VHE range. There are two main techniques used on the ground. One is imaging atmospheric Cherenkov telescope (IACT), such as H.E.S.S. \cite{Abramowski2015}, MAGIC \cite{Aleksic2012} and VERITAS \cite{Allen2017}, which observe the Cherenkov light emitted by secondary particles generated in the air showers. The other is extensive air shower (EAS) array technique, like Tibet AS$\gamma$ \cite{Amenomori2005a} and ARGO-YBJ \cite{Bartoli2014}, where secondary particles are detected at ground level. The use of the water Cherenkov technique for gamma-ray observations was developed by Milagro \cite{Becker2007}, where secondary particles (e$^{\pm}$ and muons ) go through pure water. This technique allows for better photon/hadron discrimination compared with EAS arrays. Differing from the excellent angular and energy resolution as well as the strong background-rejecting capability of IACT, the water Cherenkov detector exhibits a high duty cycle and a wide field of view (FOV) with moderate angular resolution and background rejecting ability. Therefore, it is suitable to monitor the whole sky and observe extended sources.\par
HAWC has reported the detection of at least 39 sources in a Northern sky survey \cite{Abeysekara2017}. Given the larger effective area of the WCDA, it will provide an improvement on previous and current experiments like HAWC. The prospects of the WCDA to search for $\gamma$-rays are presented in this paper. We introduce the WCDA in Section 2 and describe the properties of sources and the simulation process in Section 3. The analysis method, namely the all-sky method, is presented in Section 4. Finally, we predict the significance of the detection of the sources and the diffuse $\gamma$-rays in Section 5.\par
\section{The Water Cherenkov Detector Array}\label{sec:WCDA}
The WCDA detects showers in a primary energy range from 100 GeV to 20 TeV and constitutes one important part of LHAASO, located in Daocheng site, Sichuan province, P.R. China ($29^{\circ}21^\prime31^{\prime\prime}$ N, $100^{\circ}08^{\prime}15^{\prime\prime}$ E ), at an altitude of 4410 m. The original array covers 90,000 m$ ^2$ as reported in \cite{2009icrc}, which is divided into 4 subarrays with a size of 150 m $\times$ 150 m. Each subarray contains 900 detector units measuring 5 m $\times$ 5 m. One upward-facing 8-inch photomultiplier tube (PMTs) is anchored at the center of the unit bottom. The prospects of the WCDA reported in this paper are based on this original configuration. The design has been modified and we will discuss the effect of such difference on our result in Section 6. \par
The simulation according to the original design mentioned above was reported in \cite{Yao} \cite{Bai2019}. This simulation adopts CORSIKA6735 \cite{1998cmcc.book.....H} to simulate the cascade processes of $\gamma$-rays and cosmic rays in the atmosphere. A program based on GEANT4 \cite{2003NuPhS.125...60G} is employed to study the detector responses. This simulation tracks the Crab Nebula, a source typically used as a ``standard candle'' in VHE $\gamma$-ray astronomy, to generate $\gamma$-ray and CRs events. We name this simulation as Crab-centered simulation. We adopt the Crab spectrum measured by HEGRA \cite{Bartoli2015}, and the spectra of cosmic rays follow the H{\"o}randel model \cite{2003APh....19..193H}. Then, the directions of the simulated events are reconstructed by fitting the shower fronts. Based on this simulation, we select a data set to study some properties of the WCDA, mainly the effective area, point-spread function (PSF), and photon/hadron discrimination. There are three principles for the data selection: firstly, the reconstructed zenith angle is less than 45$^\circ$; secondly, the number of triggered detectors is more than 128; thirdly,  the ``compactness'' \cite{Hampel-Arias2015} \cite{Atkins2003}  is larger than 14.4. The compactness is defined as the $nPMTs/CxPE_{45}$, where $CxPE_{45}$ is the maximal energy deposition measured in photo-electrons (PEs) recorded by one PMT beyond a radius of 45 meters from the reconstructed air shower core; $nPMTs$ denotes the number of triggered PMTs. The median energy for this data set is around 1 TeV. After getting these properties of the WCDA, the detecting ability of WCDA to other sources is estimated from a fast simulation described in Section 3.

In detail, the effective area for $\gamma$-rays as a function of energy and zenith angle is shown in Figure \ref{FigAga}, that of CRs in Figure \ref{FigAp}. The effective area for $\gamma$-rays is used to generate the $\gamma$ signals from different sources, and the effective area for cosmic rays is used to produce the backgrounds as explained in Section 3. 

\begin{center}
\includegraphics[width=7cm]{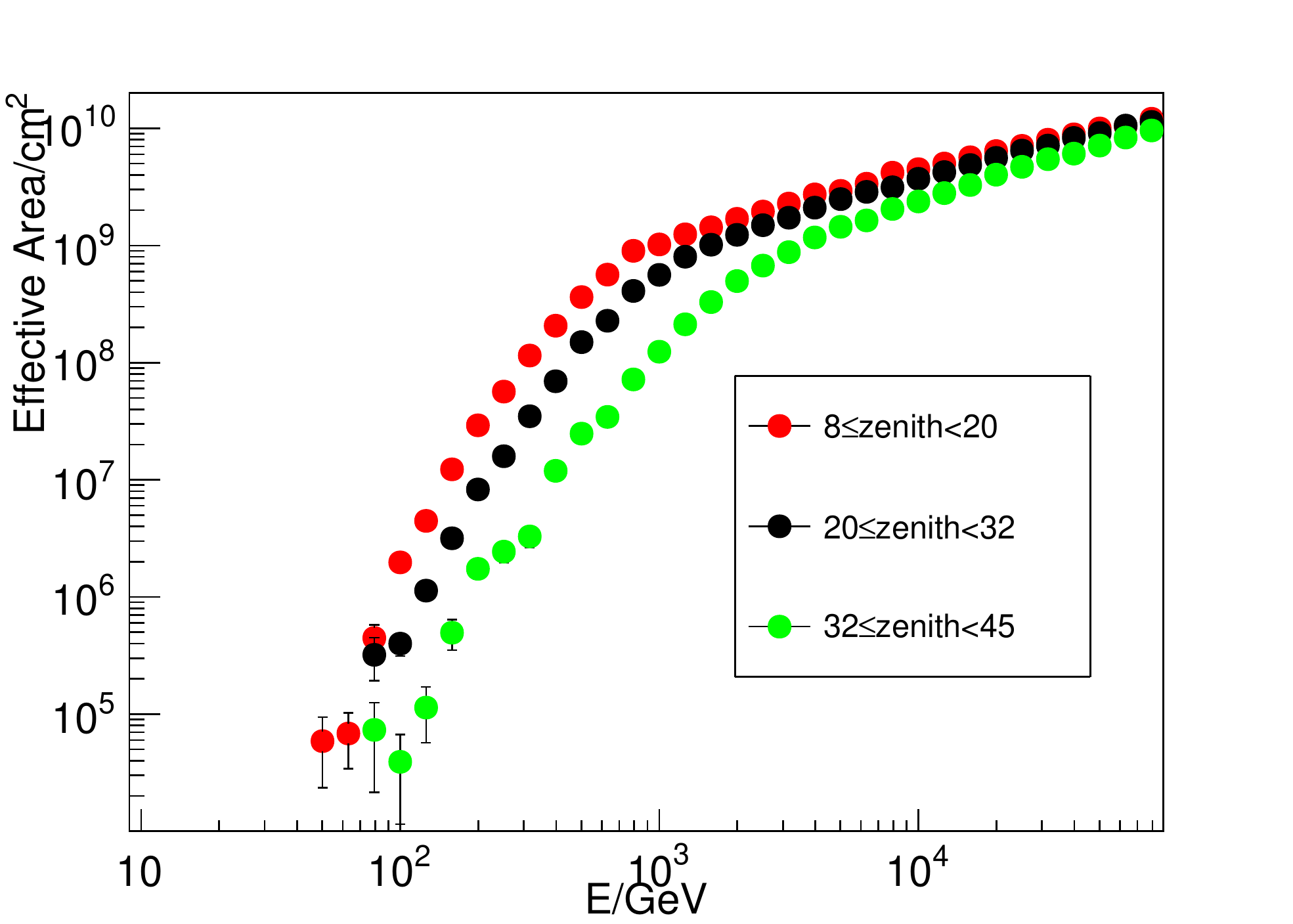}
\figcaption{\label{FigAga}The $\gamma$-ray effective area of the WCDA as a function of energy and zenith angle. Red dots denote the effective area in $8^\circ - 20^\circ$ zenith angle range; black dots denote the effective area in $20^\circ - 32^\circ$ zenith angle range; green dots denote the effective area in $32^\circ - 45^\circ$ zenith angle range. }
\end{center}
\begin{center}
\includegraphics[width=7cm]{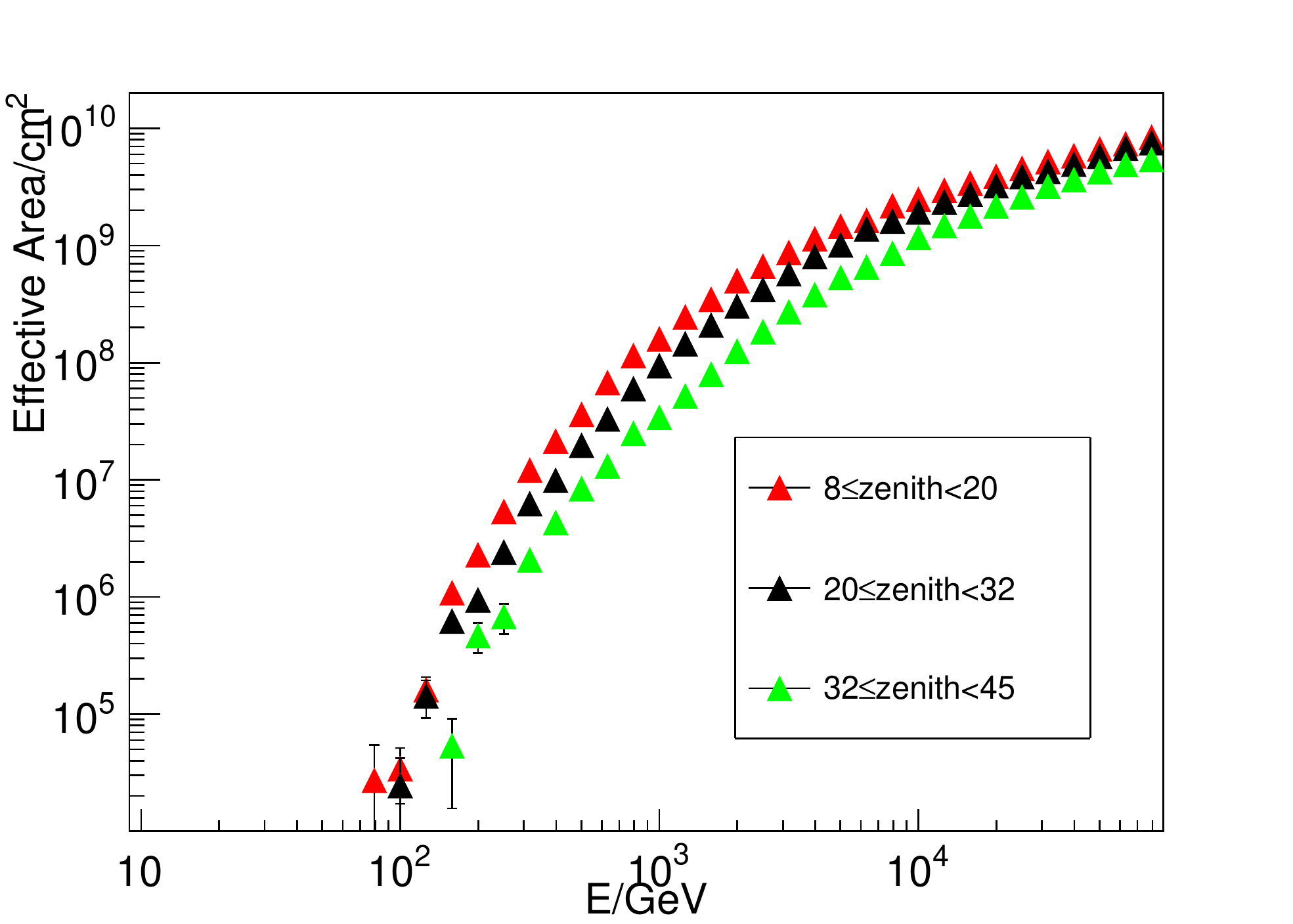}
\figcaption{\label{FigAp}The cosmic-ray effective area of the WCDA as a function of energy and zenith angle. Red dots denote the effective area in $8^\circ - 20^\circ$ zenith angle range; black dots denote the effective area in $20^\circ - 32^\circ$ zenith angle range; green dots denote the effective area in $32^\circ - 45^\circ$ zenith angle range. }
\end{center}

The PSF describes the difference between the original direction and the reconstructed direction after accounting for the detectors response. In the selected data set, the PSF is shown in Figure \ref{Fig.sub.2}, and the direction of the $\gamma$-ray signals from sources is smeared with this function. The blue dashed line in Figure \ref{Fig.sub.2} shows the PSF of the WCDA, and the red line shows the PSF in our fast simulation. The two lines agree well which proves that our fast simulation spread the signals from $\gamma$ sources properly according the PSF of the WCDA. PSF convolves with energy. We analyze only one energy bin with the median energy of 1 TeV, so one specific PSF is used. Due to the PSF of the WCDA, $\gamma$-ray signals from a point source follow a central symmetric distribution around the source. We integrate signal and cosmic-ray background counts within a circular disc centered on a sky position. The optimal disc radius is found by maximizing the figure of merit $S/\sqrt{B}$ where $S$ is the number of signal counts, and $B$ of background counts. The best signal-to-noise ratio occurs at 0.56$^\circ$, denoted as the red vertical line in Figure \ref{Fig.sub.2}. Therefore, the angular smoothing radius is 0.56$^\circ$.
\begin{center}
\includegraphics[width=7cm]{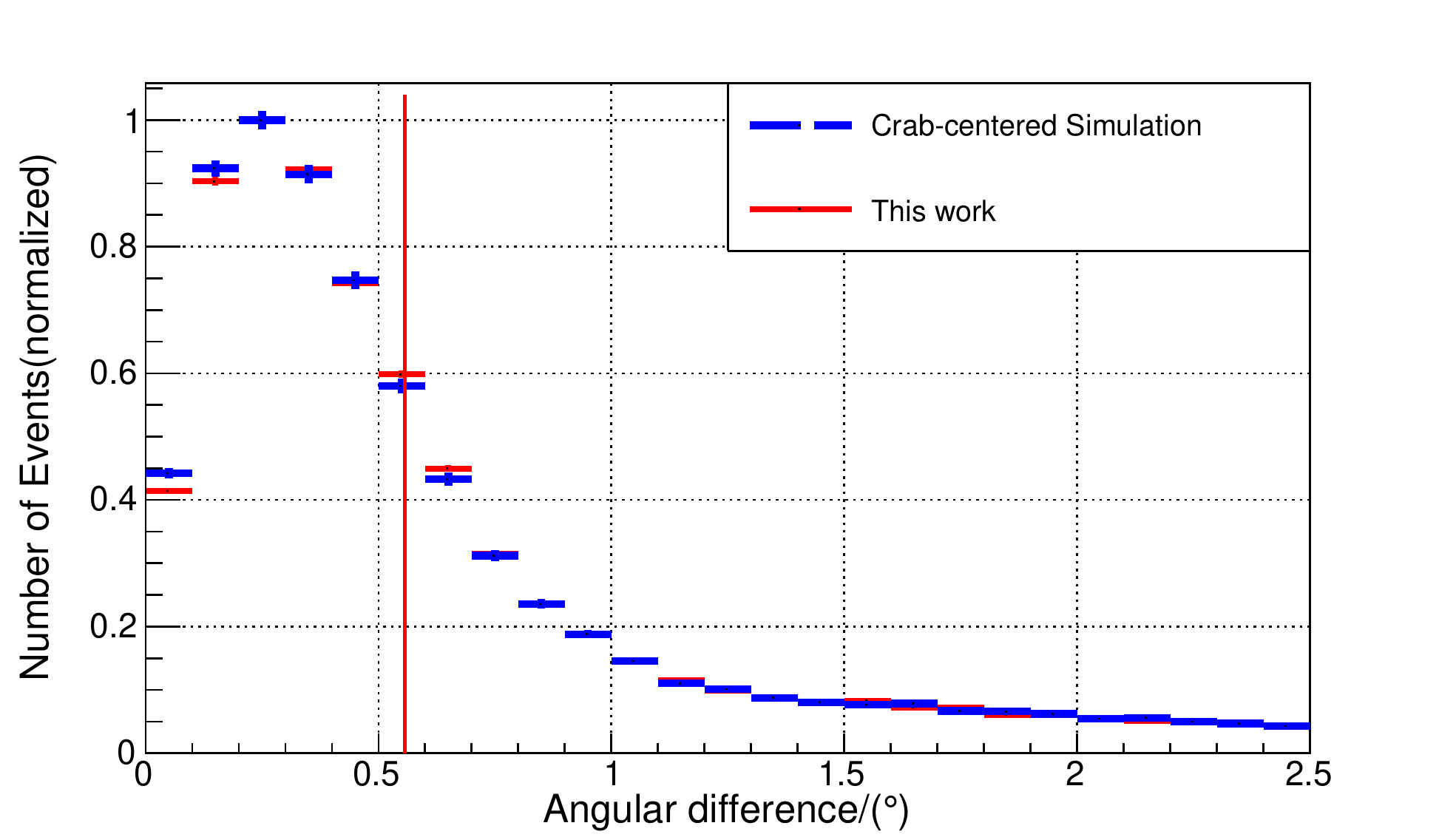}
\figcaption{\label{Fig.sub.2}The PSF for WCDA. The blue dashed line is the PSF derived from the selected data set. The red solid line is the PSF in our fast simulation. The two lines agree well which ensures the reliability of our fast simulation. The red vertical line denotes the optimized angular radius(0.56$^\circ$) with the best signal-to-noise ratio.}
\end{center}

The WCDA adopts the parameter named compactness to discriminate $\gamma$-rays from cosmic rays. Statistically, the compactness distributions of $\gamma$-rays and cosmic rays are different as shown in Figure \ref{Fig.cmpt}. The compactness of $\gamma$-ray showers is smaller than that of cosmic rays, because secondary muons are more likely to deposit energy in PMTs far from the air shower core, and they mainly originate from hadronic cosmic rays interactions with the atmosphere. We quantify the performance of the photon/hadron rejection method by calculating its \textit{Q}-factor as defined: $ Q=\frac{\eta_{\gamma}}{\sqrt{\eta_{cr}}}$, where $\eta_{\gamma}$ and $\eta_{cr}$ are the efficiency to keep the simulated $\gamma$-rays and cosmic rays when the compactness is greater than a value. We scan for the maximal \textit{Q}-factor value by varying the compactness values as shown in Figure \ref{Fig.cmptQ}. For this set of data, the optimized photon/hadron discrimination criterion is compactness $>$ 14.4, where the efficiency of $\gamma$-rays ($\eta_{\gamma}$) and CRs ($\eta_{cr}$) is 40\% and 0.27\% respectively.
\begin{center}
\includegraphics[width=7cm]{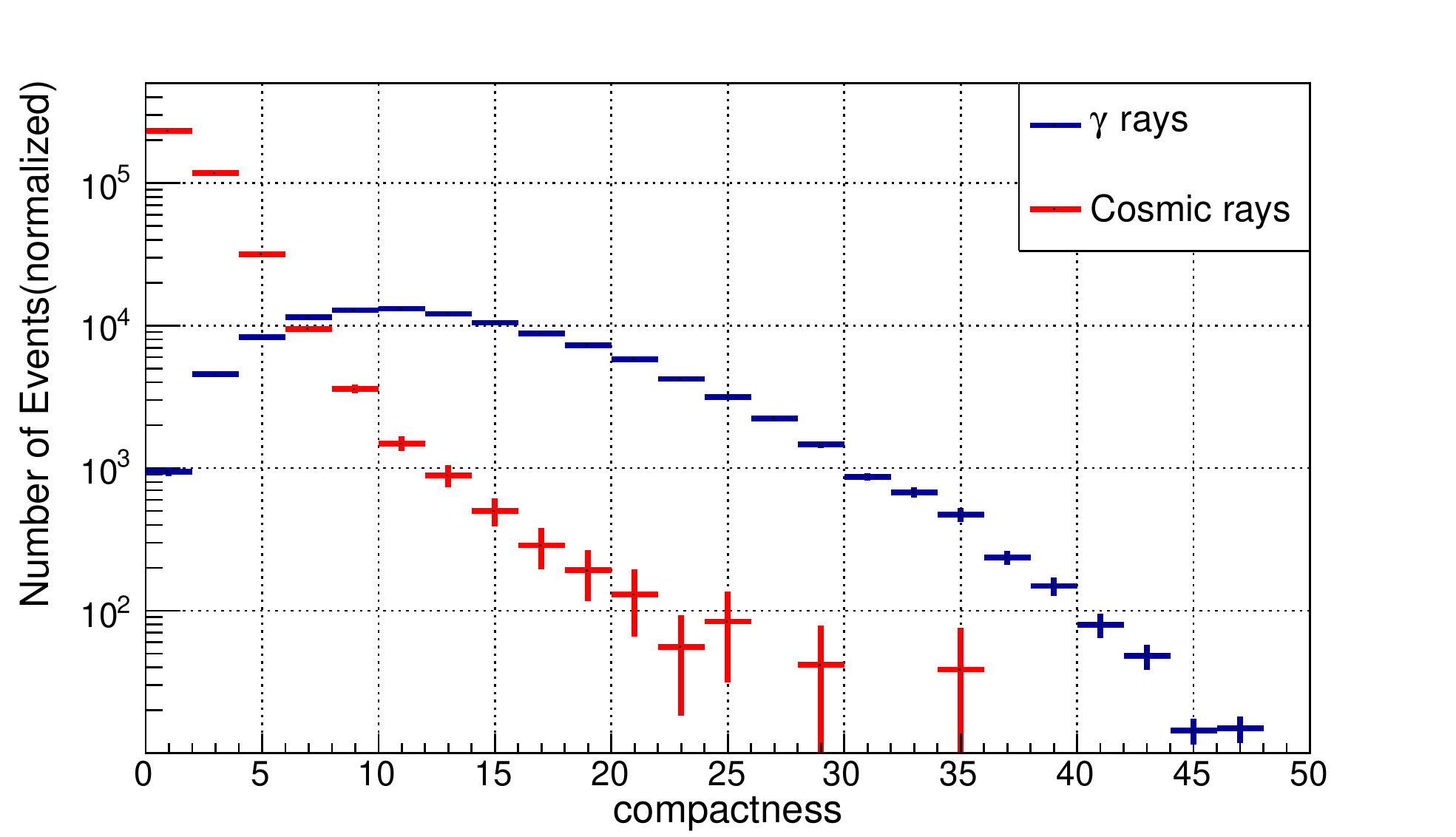}
\figcaption{\label{Fig.cmpt}The compactness distribution of $\gamma$-rays (red line) and cosmic rays (blue line) separately.  }
\end{center}
\begin{center}
\includegraphics[width=7cm]{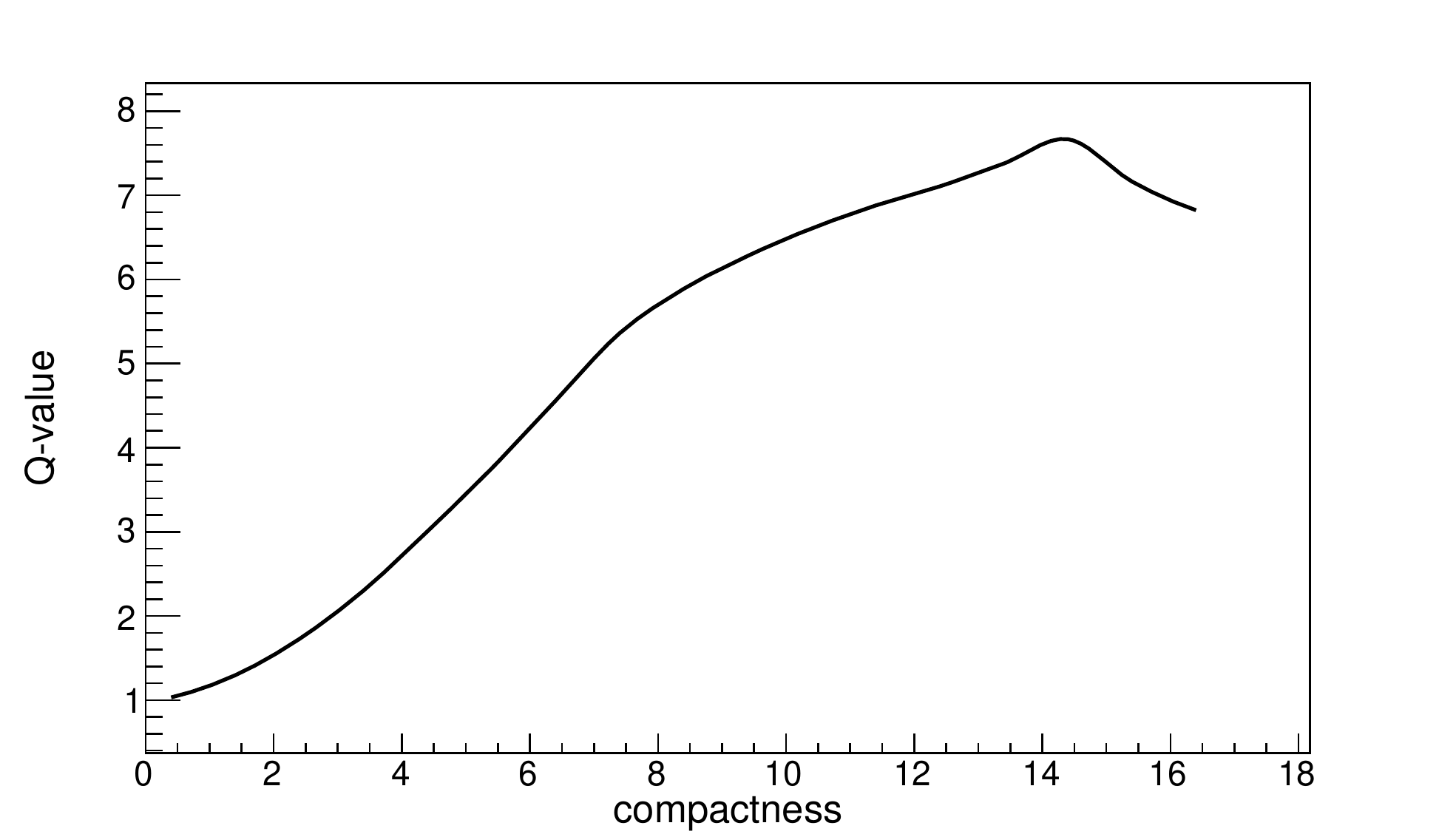}
\figcaption{\label{Fig.cmptQ}The \textit{Q}-factor value as a function of the compactness value.} 
\end{center}

\section{ Fast Simulation}\label{sec:FS}
Separately from the Crab-centered simulation, we perform a fast simulation of the array exposure across its field of view (FOV) to calculate the detection significance of all sources in TeVCat. In this work, the FOV of WCDA is defined as the portion of the sky with a zenith angle $\leq 45^{\circ}$. We project the FOV in local coordinates (zenith and azimuth) , in which the zenith angle ($\theta$) is binned in $0.08^{\circ}$-angle bins and the azimuth ($\phi$) is binned in $\frac{0.08^{\circ}}{\sin\theta}$-angle bins, so that each window contains the same steradian units for solid angle $\Omega=1.95\times10^{-6}$. At the same time, a sidereal day is divided into 3600 time bins, in other words, one day contains 3600 maps with an exposure time of 24 seconds.
The predicted number of cosmic rays or diffuse $\gamma$ rays in a window $(t,\theta,\phi)$ is calculated as
\begin{eqnarray}
\label{con:Acr}N_{i}(t,\theta,\phi)=\eta_{i}\int_{E} \phi_{i}(E)A_{i}(\theta,E)\Omega\, dE\delta t 
\end{eqnarray}
where $\Omega$ is in steradian unit for solid angle; $\delta t$ is the period of one map that is 24 seconds.
When $i$ denotes the CRs, $A_{i}(\theta,E)$ is the differential effective area of cosmic rays; $\phi_{i}(E)$ is the cosmic ray spectrum \cite{Gaisser2013}; $\eta_{i}$ is the efficiency of CRs which passed the photon/hadron criterion. When $i$ denotes the diffuse $\gamma$-rays, these parameters are values of $\gamma$ rays. The diffuse $\gamma$-rays spectra are the results from the paper \cite{Guo2016}.
We track every source located in the FOV and calculate the number of $\gamma$-ray events from each source. The predicted number in a window $(t,\theta,\phi)$ is calculated as
\begin{equation}
\label{con:Aga}N_{\gamma}(t,\theta,\phi)=\eta_{\gamma}\int_{E} \phi(E)_{\gamma}A_{\gamma}(\theta,E)\, dE\delta t
\end{equation}
The meaning of each parameter is the same as (\ref{con:Acr}) but represents the property of $\gamma$-ray, excluding the solid angle. The spectra of the sources that we use are listed in the Table \ref{table:SNRs}, \ref{table:PWNe}, \ref{table:AGN}, \ref{table:UID}. The spectra of sources consist of a power law with a fixed index: $\phi(E)=N_{0}(\frac{E}{E_{0}})^{-\beta}$, where $N_0$ is the differential flux at $E_0$, and $\beta$ is the spectral index. If spectra of sources are measured with an exponential energy cut, the spectra are in the form $\phi(E)=(\frac{E}{E_{0}})^{-\beta}e^\frac{-E}{E_{cut}}$, where $E_{cut}$ is the exponential cutoff energy of sources. If sources are extended sources, the extension is determined by fitting to the excess map with a two-dimensional (2D) Gaussian convolved with the PSF \cite{Aharonian2008}. Therefore, we use the 2D Gaussian model to produce the morphologies of extended sources. The used parameters for each source are listed in Table \ref{table:SNRs},\ref{table:PWNe},\ref{table:AGN},\ref{table:UID}.

\section{Analysis Method}\label{sec:Method}
Since events in each pixel contain both $\gamma$-ray signals and background CRs, the key point is to estimate the number of background properly and test whether there is a significant excess. We use the All-Sky analysis method to estimate the background events, which has been already successfully used in Tibet AS$\gamma$ experiment \cite{Amenomori2005}. \par
The detection efficiency largely depends on the zenith angle, because more inclined events will go through a greater atmospheric depth. However, the efficiency in one zenith belt is independent of azimuth angle, given that the WCDA is almost sitting on a horizontal plane. When we estimate the background events of one window in the fast simulation, this window is called an ``on-source window" and the sideband windows in the same zenith angle belt are usually referred to as ``off-source window". The background events of ``on-source window" is estimated by the average number of ``off-source window"s. The FOV in equatorial coordinates is divided into small pixels measured  $0.1^\circ \times 0.1^\circ$, and each window marked as $(t,\theta,\phi)$ in the fast simulation corresponds to a pixel marked as $(i,j)$ in equatorial coordinates. We denote the number of events in on-source window as $ N_{t,\theta,\phi}$  and the relative intensity as $I_{i,j}$, the number of events in the $\phi'$-th off-source window as $ N_{t,\theta,\phi'}$ and the relative intensity as $ I_{i',j'}$. We can derive:$\frac{N_{t,\theta,\phi}}{I_{i,j}}=<\frac{N_{t,\theta,\phi'}}{I_{i',j'}}>$. For the FOV of WCDA,
\begin{equation}
\label{con:chi2}
\tilde{\chi}^2=\sum_{i,j}^{}\left(\frac{
\frac{N_{t,\theta,\phi}}{I_{ij}} -\frac{1}{n_{\theta}-1}
\sum_{\phi'}\frac{N_{t,\theta,\phi'}}{I_{i'j'}} 
}{{\sigma_{t,\theta,\phi}}}\right)^2
\end{equation}
Where $n_{\theta}$ represents the number of windows in $\theta$-zenith belt.
We will get the relative intensity $I_{i,j}$ and the estimated error $\delta I_{i,j}$ by minimizing the $\tilde{\chi}^2$. The background of each pixel is $N_{bkg i,j} = \frac{N_{i,j}}{I_{i,j}}$. The relative intensity gives the amplitude of deviations in the number of events  from the backgrounds expectations. The significance of deviations can be calculated as $\sigma =\frac{I_{i,j}-1}{\delta I_{i,j}}$. 

In the fast simulation, the skymap contains $\gamma$-rays from both the sources and the diffuse emissions. However, the signal counts from sources near the Galactic plane may have underlying diffuse component. We adopt the likelihood ratio method to decompose the two components \cite{Abeysekara2017}.$$\mathcal{L} = \ln \frac{\mathcal{L}(signal\quad model)}{\mathcal{L}(Null{\quad}model)}$$ In the following analysis, the signal model only considers the signal counts from two components: $M_{i,j}=N^{'}_{i,j} + N^{'}_f$. $N^{'}_{i,j}$ is the source contribution to the pixel $(i,j)$ and derived from the source flux and the detector response. The morphologies of the point sources are described by the PSF and those of extended sources can be characterized by the extended source shapes (2D Gaussian model) convolved with the PSF. To evaluate the maximum possible contribution of the diffuse emission to source signal counts, we assume that $N^{'}_f$ is a constant number for each pixel in a circular $3^\circ$ region of interest (ROI) centered on our source. Therefore, the signal likelihood follows $\mathcal{L}(signal\quad model) = \sum_{i,j} \ln P_{i,j}(N_{i,j},N_{bkg i,j}+M_{i,j})$, where $P_{i,j}$ is the Poisson probability of observing $N_{i,j}$ counts given the expectation $N_{bkg i,j}+M_{i,j}$. As for the null model, the expectation only considers background counts $N_{bkg i,j}$. We use the minuit library \cite{1975CoPhC..10..343J} to maximize the likelihood ratio.
\section{Results }
The sensitivity of the WCDA with declination is presented in Figure \ref{FigSd} and the spectrum index is -2.62.
\begin{center}
\includegraphics[width=7cm]{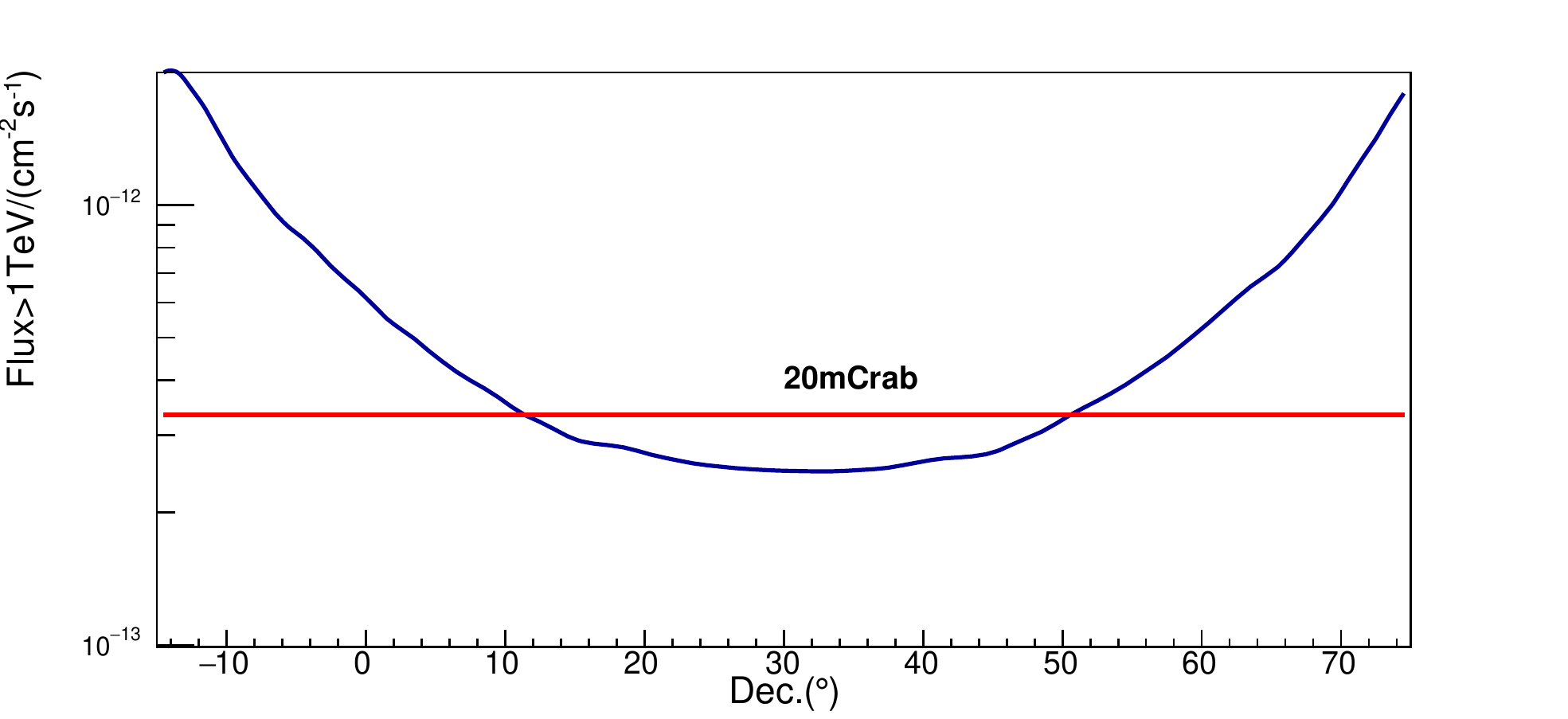}
\figcaption{\label{FigSd}  Sensitivity changes across declinations.}
\end{center}

The $\gamma$-ray signals from sources are based on the spectra in TeVCat and the spectra of diffuse emissions is calculated by the spatially-dependent diffusion model \cite{Guo2016}, which well accounts for the Galactic plane flux measured by \textit{Fermi}-LAT \cite{Ackermann_2012}. Therefore, we use this model to calculate the spectra of diffuse $\gamma$-rays in the TeV range. In this model, the diffuse volume is contributed by two regions, one is close to the Galactic disk which is called the inner halo, and the other is the outer halo. The turbulences in the inner halo are originated from supernova explosions, while in the outer halo, the turbulences are mainly generated by CRs themselves. Therefore, the energy spectra of turbulences in the inner halo are harder than that in the outer halo while the diffusion is slower. In the outer halo, the diffusion coefficient is only rigidity dependent, while that in the inner halo is both rigidity and spatially-dependent, which is anticorrelated with the SNR distribution. We used DRAGON code \cite{Evoli_2008} to numerically solve the distribution of CRs. The CRs interact with interstellar medium of the Milky Way to produce diffuse $\gamma$-rays. The average $\gamma$-ray flux of the inner Galactic plane $(22^\circ \leq l \leq 62^\circ, -2^\circ \leq b \leq 2^\circ )$ is shown in Figure \ref{Fig.diff_spec}. The black line shows the total diffuse emission of three processes, the dashed green line shows the $\gamma$-rays produced via $\pi^{0}$ decay, the dashed blue and green lines denote the $\gamma$-rays produced by electrons via inverse Compton (IC) and bremsstrahlung process, respectively. The exponential cutoff at tens of TeV is due to the cutoff of the injection spectrum at 150 TeV, which is corresponding to the CRs' spectrum measured by CREAM \cite{2018ChPhC..42g5103G}. Actually, the cutoff energy of diffuse  $\gamma$-rays is one magnitude higher than the median energy in our work, and it will not affect our results.
\begin{center}
\includegraphics[width=7cm]{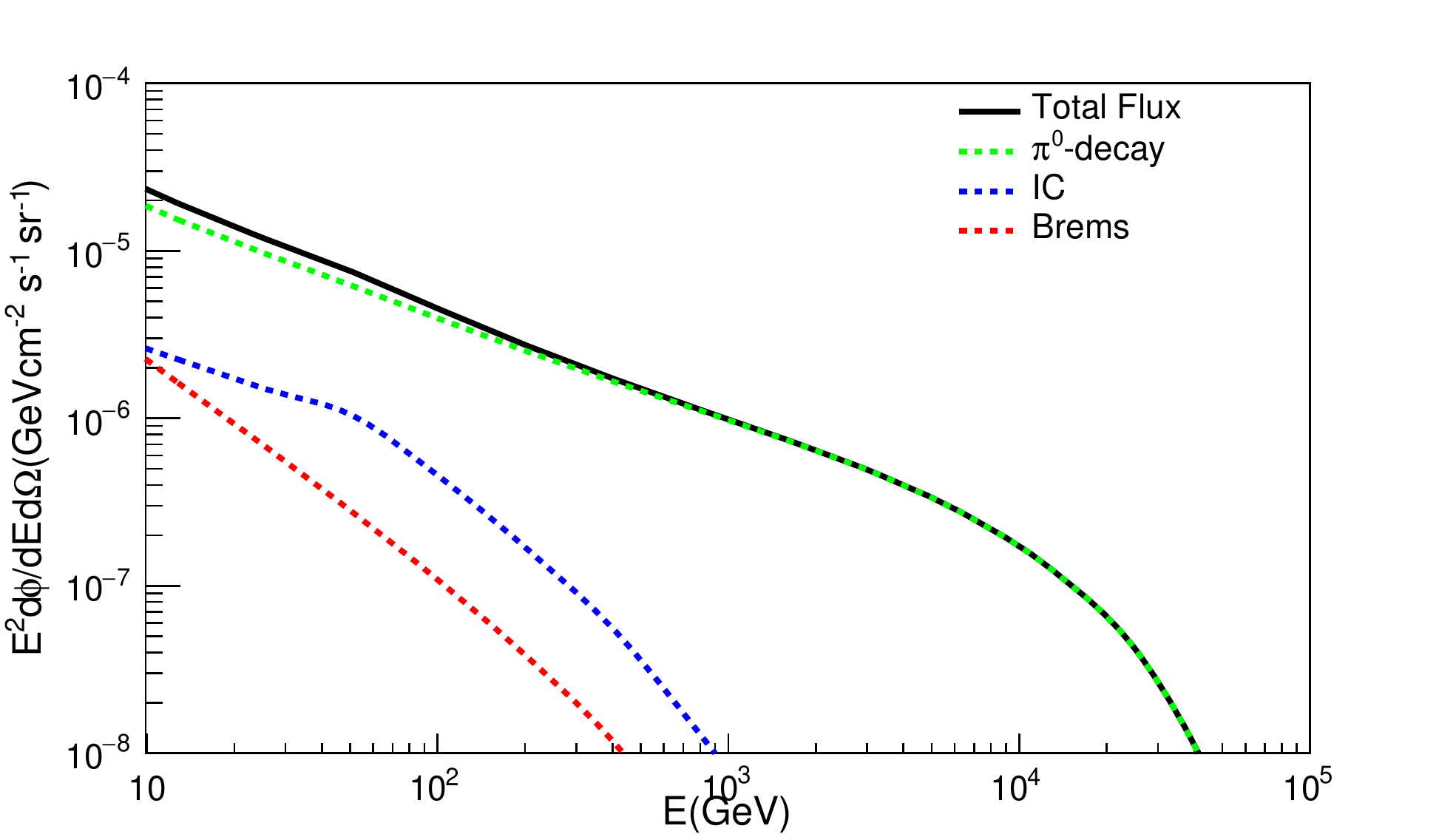}
\figcaption{ \label{Fig.diff_spec}The average flux of the inner Galactic plane ($22^\circ \leq l \leq 62^\circ, -2^\circ \leq b \leq 2^\circ$, $l$ is the Galactic longitude and $b$ is the Galactic latitude). The black line shows the total diffuse emission of three processes, the dashed green line shows the $\gamma$-rays produced via $\pi^{0}$ decay, the dashed blue and green lines denote the $\gamma$-rays produced by electrons via inverse Compton (IC) and bremsstrahlung process.}
\end{center}

We generate two skymaps, one is the map that only considers the TeV sources, and the other is the map that only demonstrates the diffuse emissions. Then we combine the two skymaps to analyze the prospect of their detections with one-year exposure. The one-dimensional projection of significance is shown in Figure \ref{Fig.sig_one}. The red line is a standard normal distribution and the black line is the significance distribution across the sky. For lower values, the significance is well reproduced by the normal distribution. However, the greater values are due to the $\gamma$-ray sources and diffuse emission. The two-dimensional skymap is shown in Figure \ref{Fig.all-sky}. Although the significances of many of sources are greater than 15, Figure \ref{Fig.all-sky} limits the range of significance from -5 to 15 for visualization. 
\begin{center}
\includegraphics[width=7cm]{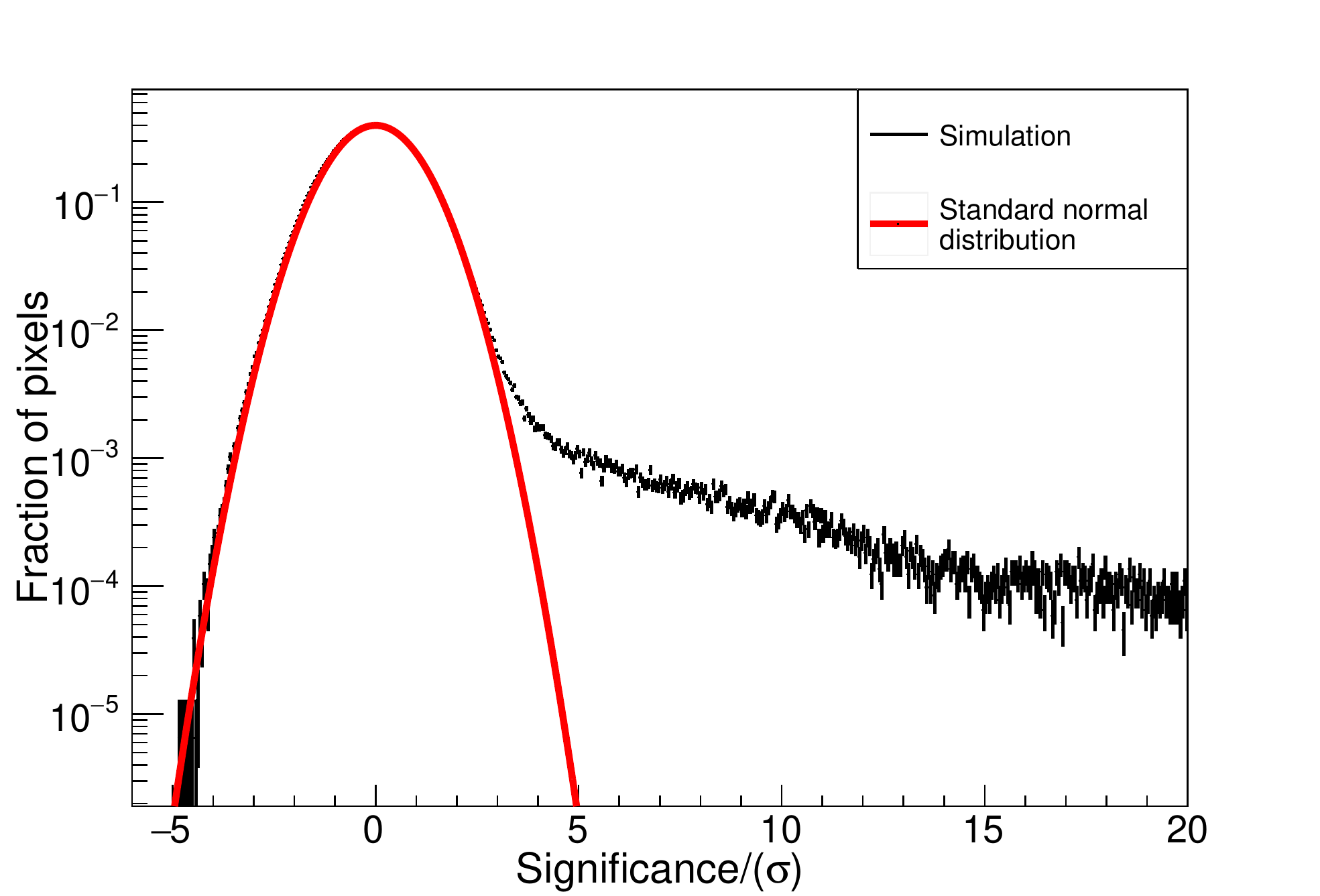}
\figcaption{\label{Fig.sig_one}The significance distribution for the sky map (black line) and a standard normal distribution (red line).}
\end{center}
\end{multicols}
\begin{center}
\includegraphics[width=14cm]{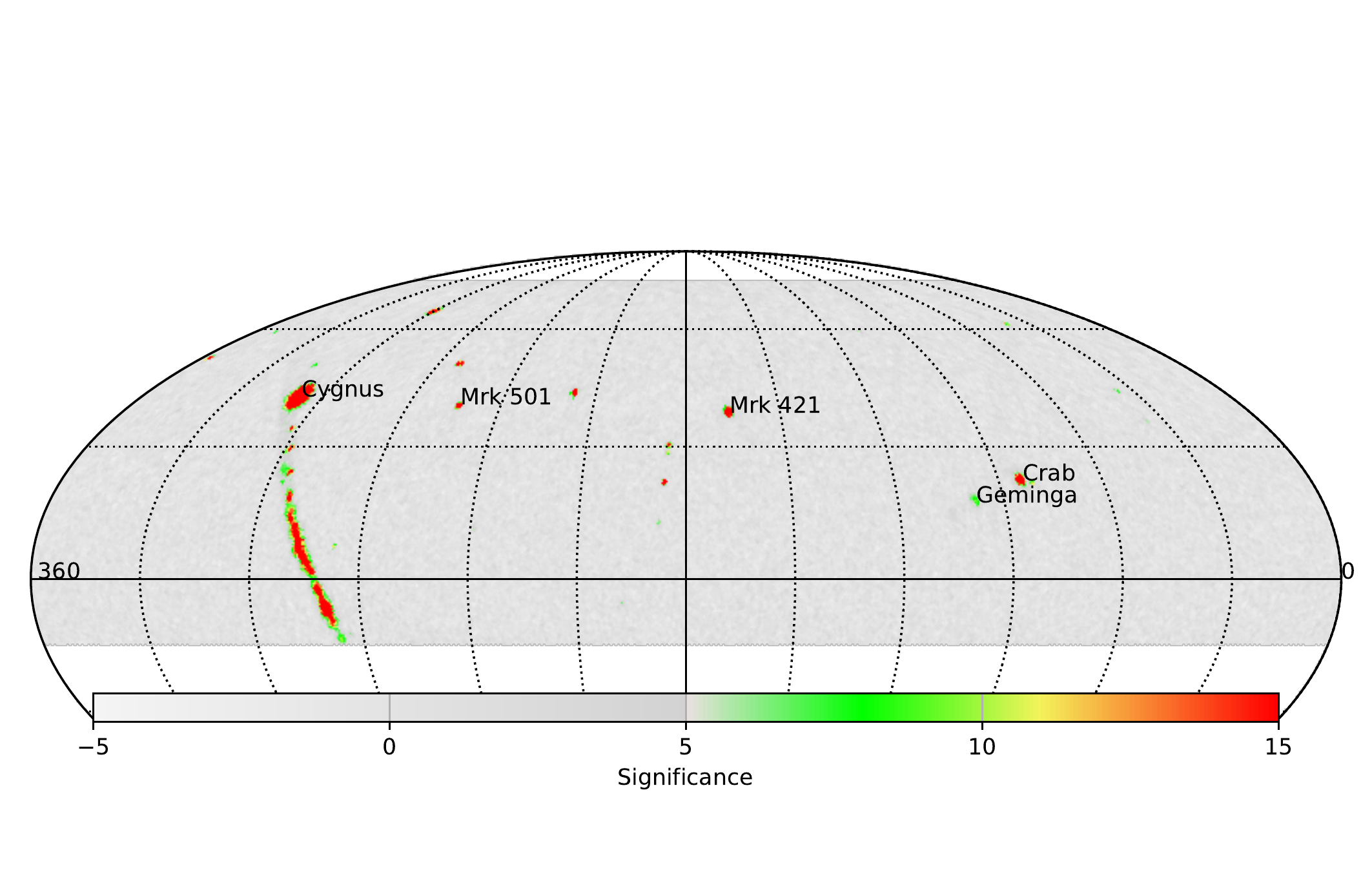}
\figcaption{\label{Fig.all-sky}The significance of all TeV sources and diffuse emission in equatorial coordinates (J2000.0 epoch), we limited the significance from -5 to 15 for visualization. }
\end{center}
\begin{multicols}{2}
The combined skymap presented in Figure \ref{Fig.all-sky} includes both TeV source signals and diffuse emission. Actually, there are 22 ( ARGO J2031+4157, LS I +61303, HESS J1912+101, W51, HESS J1831-098, 2HWC J1837-065, 2HWC J1825-134, MAGIC J1857.6+0297, TeV J1930+188, 2HWC J1844-032, 2HWC J1852+013, HESS J1858+020, 2HWC J2006+341, 2HWC J1902+048, 2HWC J1907+084, MGRO J1908+06, 2HWC J1914+117, 2HWC J1921+131,2HWC J1928+177, 2HWC J1938+238, 2HWC J1953+294, 2HWC J1955+285) sources on the Galactic plane ($-2^\circ \leq b \leq 2^\circ$) and the signals from these sources may be overestimated due to the diffuse $\gamma$-ray contributions in the combined skymap. We decompose the two components as described in Section 4 and calculate the signal counts ($N_{calcu}$) of these sources from the combined map. To estimate the uncertainties caused by diffuse emission in our analysis, we compare the $N_{calcu}$ of sources in the combined skymap with the signal counts ($N_{detected}$) in the source map, because $N_{detected}$ is the source signal counts, considering that the source skymap excludes diffuse emission. The ratio of $N_{calcu}$ to $N_{detected}$ changing along $N_{detected}$ is shown in Figure \ref{Fig.de}, the $N_{calcu}$ in the combined map tend to $N_{detected}$, especially when the source signal counts are large, and we can limit the uncertainty of the diffuse emission to the level of 20\%, which agrees with results in \cite{Abeysekara2017}. After subtracting the diffuse emission, the predicted significances and detailed information of sources (location, spectrum, energy cutoff, extension) are presented in Table \ref{table:SNRs},\ref{table:PWNe},\ref{table:AGN},\ref{table:UID}. Among the observed sources with significances greater than 5$\sigma$, there are 29 Galactic sources, constituting 20 unidentified sources, 4 PWN, and 5 other sources (superbubbles, SNRs, Shells and Binaries). There are 13 extragalactic sources, all of which are AGN. Another work \cite{Zhao2016} predicts that 9 sources (Mrk 421, 1ES 1215+303, 1ES 1218+304, W Comae, H 1426+428, 1ES 1959+650, Mrk 501, 1ES 2344+514, RGB J0710+591) will be detected after considering extragalactic background light (EBL) absorption effect, which agrees with our results. We also predict the WCDA will detect M 87 with 6.82 $\sigma$, and this source is not included in the work \cite{Zhao2016}. Beyond these sources, the spectra of S3 0218+35 and RGB J2056+496 are measured in flare sates; the redshift of VER J0521+211 is more than 0.1, while we adopt an extrapolated spectrum from observation and do not consider the EBL absorption effect.
\begin{center}
\includegraphics[width=7cm] {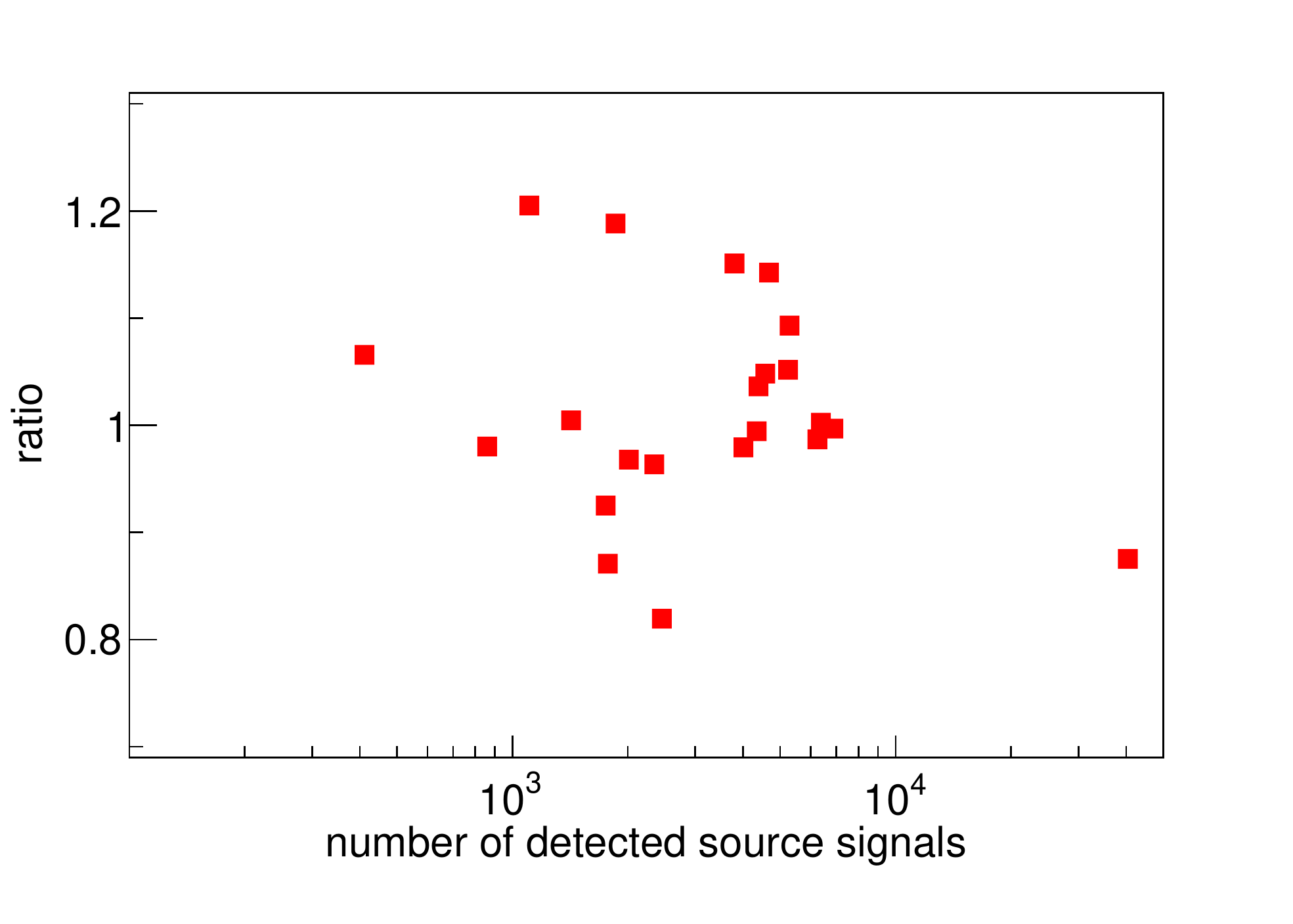}
\figcaption{\label{Fig.de}The uncertainty of the source signal counts on the Galactic plane ($-2^\circ \leq b \leq 2^\circ$) caused by the diffuse emission in my analysis. The Y-axis is the $N_{calcu}$ to $N_{detected}$ ratio, and the X-axis is $N_{detected}$ in the source skymap.}
\end{center}
\section{Discussion }\label{sec:discussion}
The WCDA is designed to detect $\gamma$-rays from hundreds of GeV to tens of TeV to study the propagation and acceleration of cosmic rays. The results obtained in this work unveil the scientific potential of WCDA to search for $\gamma$-ray sources. We study the $\gamma$-ray sources and diffuse emissions simultaneously and then figure out the sources that have the potential to be observed with a significance more than 5$\sigma$ within WCDA's one-year exposure.

The ground-based IACTs have detected tens of AGN at VHE. Compared to IACTs, the WCDA has a wide FOV and long duty time, which makes it potential to detect AGN with long-term emissions. The prospects of the WCDA to detect the already-known AGN are presented in Table \ref{table:AGN}. However, there are two uncertainties in these results. One is the unpredictable variability on AGN flux. We adopt the measured time-averaged spectra prior to spectra in flare states and extend the spectra without energy cutoff. At the same time, we assume that these sources have constant flux and then calculate the significance of their detection. The other is the absorption effect by extragalactic background light. We are limited to these nearby AGN (the redshift distances of these AGN are less than 0.13 except S3 0218+35) whose corresponding optical depth is less than 1, and that means the $\gamma$-rays emitted by these AGN will not be strongly absorbed. Therefore, this work ignores the EBL absorption effect. 

As we mentioned in Section 2, the design of WCDA has been modified, and the new design is described in \cite{Bai2019}. The area is changed from 90,000 $m^2$ to 7,8000 $m^2$ because one large subarray measuring 300 m $\times$ 110 m will replace two original subarrays. The number of detector units is 3210, 390 fewer than the original design. A reduction in the effective area results in a sensitivity reduction of approximately 20\%. Moreover, one detector unit consists of two PMTs at the center of each cell's bottom. In the first pond, each detector unit consists of one 8-inch and one 1.5-inch PMTs, while each detector unit in the other two ponds consists of one 20-inch and one 3-inch PMTs. The small PMTs would work as a joint observation with Cherenkov telescope array (WFCTA) above 100 TeV, and the change from 8-inch PMTs to 20-inch PMTs aims to improve the sensitivity around 100 GeV. Since our analysis is performed at 1 TeV, such differences would not change our results significantly.\par
\end{multicols}
\begin{center}
\tabcaption{ \label{table:SNRs} Significance of superbubbles, SNRs, Shells, Binaries. $\sigma$ is the significance of sources, $N_0$ is the differential flux at $E_0$, $\beta$ is the spectral index, extension is the extended angular radius in degrees under the assumption of two-dimensional Gaussian model.}
\footnotesize
\begin{tabular*}{170mm}{@{\extracolsep{\fill}}lcccccccc}
\toprule $TeVCat Name$ & $R.A./^\circ$   & $Dec./^\circ$ & $\sigma$ &  $N_0/(TeV^{-1 }cm^{-2}s^{-1})$ &  $E_{0}/TeV$ & $\beta$ & $Extension/^\circ$ & Ref. \\
\hline 
LS I +61303 & 40.14 & 61.26 & 9.4 & $1.80\times 10^{-12}$ & 1 & 2.34 & \_\_ &\cite{Archambault2016}\\
HESS J1912+101 & 288.20 & 10.15 & 9.7 & $3.66\times 10^{-14}$ & 7 & 2.64 & 0.7 & \cite{Abeysekara2017} \\
W51 & 290.73 & 14.19 & 10.0 & $2.61\times 10^{-14}$ & 7 & 2.51 & 0.9 & \cite{Abeysekara2017} \\
ARGO J2031+4157$^a$ & 307.8 & 42.50 & 67.5 & $3.50\times 10^{-9}$ & 0.1 & 2.16 & 2 & \cite{Bartoli2014} \\
Cassiopeia A & 350.81 & 58.81 & 7.2 & $1.45\times 10^{-12}$ & 1 & 2.75 & \_\_ & \cite{Kumar2015} \\
\bottomrule
\end{tabular*}
\footnotesize{\\$a$: It is identified as the counterpart of the Cygnus Cocoon at TeV energies and its spectrum exhibits a exponential cutoff at the energy of 40 TeV.}
\end{center}

\begin{center}
\tabcaption{ Significance of PWN. $\sigma$ is the significance of sources, $N_0$ is the differential flux at $E_0$, $\beta$ is the spectral index, extension is the extended angular radius in degrees under the assumption of two-dimensional Gaussian model.\label{table:PWNe}}
\footnotesize
\begin{tabular*}{170mm}{@{\extracolsep{\fill}}lcccccccc}
\toprule $TeVCat Name$ & $R.A./^\circ$   & $Dec./^\circ$ & $\sigma$ & $N_0/(TeV^{-1 }cm^{-2}s^{-1})$ &  $E_{0}/TeV$ & $\beta$ & $Extension/^\circ$ & Ref. \\
\hline 
Crab & 83.63 & 22.01 & 307.7 & $1.85\times 10^{-13}$ & 7 & 2.58 &\_\_ & \cite{Abeysekara2017} \\
Geminga & 98.12 & 17.37 & 10.7 & $4.87\times 10^{-14}$ & 7 & 2.23 & 2 & \cite{Abeysekara2017} \\
HESS J1831-098 & 277.85 & -9.90 & 9.4 & $9.58\times 10^{-14}$ & 7 & 2.64 & 0.9 & \cite{Abeysekara2017} \\  
TeV J1930+188 & 292.63 & 18.87 & 23.9 & $9.80\times 10^{-15}$ & 7 & 2.74 & \_\_& \cite{Abeysekara2017} \\
\bottomrule
\end{tabular*}%
\end{center}
\begin{center}
\tabcaption{Significance of AGN. $\sigma$ is the significance of sources, $N_0$ is the differential flux at $E_0$, $\beta$ is the spectral index, $E_{cut}$ is the exponential cutoff energy of sources.
 \label{table:AGN} 
}
\footnotesize
\begin{tabular*}{170mm}{@{\extracolsep{\fill}}lcccccccc}
\toprule $TeVCat Name$ & $R.A./^\circ$   & $Dec./^\circ$ & $\sigma$ & $N_0/(TeV^{-1 }cm^{-2}s^{-1})$ &  $E_{0}/TeV$ & $\beta$ &$E_{cut}/TeV$ & Ref. \\
\hline 
S3 0218+35$^f$ & 35.27 & 35.94 & 6.4 & $2.00\times 10^{-9}$ & 0.1 & 3.8 & \_\_ & \cite{Ahnen2016} \\
VER J0521+211 & 80.44 & 21.21 & 12.4 & $1.99\times 10^{-11}$ & 0.4 & 3.44 & \_\_ & \cite{Archambault2013} \\ 
RGB J0710+591 & 107.61 & 59.15 & 5.8 & $9.20\times 10^{-13}$ & 1 & 2.69 & \_\_ & \cite{Acciari2010} \\
Markarian 421 & 166.08 & 38.19 & 236.8 & $2.82\times 10^{-11}$ & 1 & 2.21 & 5.4 & \cite{Abeysekara2017} \\
1ES 1215+303$^f$ & 184.45 & 30.10 & 5.2 & $2.30\times 10^{-11}$ & 0.3 & 3.6 & \_\_ & \cite{Aliu2013} \\
1ES 1218+304 & 185.36 & 30.19 & 17.9 & $1.40\times 10^{-12}$ & 1 & 3.13 & \_\_ & \cite{Ollaboration2013} \\
W Comae & 185.38 & 28.23 & 11.1 & $2.00\times 10^{-11}$ & 0.4 & 3.81 & \_\_ & \cite{Aliu2008} \\
M 87 & 187.70 & 12.40 & 6.8 & $7.70\times 10^{-12}$ & 0.3 & 2.21 & \_\_ & \cite{Aleksic2012} \\  
H 1426+428 & 217.14 & 42.67 & 61.6 & $4.37\times 10^{-12}$ & 1 & 3.54 & \_\_ & \cite{Petry2002} \\
Markarian 501 & 253.47 & 39.76 & 47.5 & $4.40\times 10^{-12}$ & 1 & 1.6 & 5.7 & \cite{Abeysekara2017} \\
1ES 1959+650 & 300.00 & 65.15 & 28.8 & $6.12\times 10^{-12}$ & 1 & 2.54 & \_\_ & \cite{Aliu2013yy} \\
RGB J2056+496$^f$ & 314.18 & 49.67 & 9.7 & $1.15\times 10^{-11}$ & 0.4 & 2.77 & \_\_ & \cite{Benbow2017} \\
1ES 2344+514 & 356.77 & 51.71 & 19.2 & $2.65\times 10^{-12}$ & 0.91 & 2.46 & \_\_ & \cite{Allen2017} \\
\bottomrule
\end{tabular*}
\footnotesize{\\$f$: The spectrum of this source is in a flare state.}
\end{center}

\begin{center}
\tabcaption{ Significance of unidentified sources(UID). $\sigma$ is the significance of sources, $N_0$ is the differential flux at $E_0$, $\beta$ is the spectral index, extension is the extended angular radius in degrees under the assumption of two-dimensional Gaussian model.
\label{table:UID}}
\footnotesize
\begin{tabular*}{170mm}{@{\extracolsep{\fill}}lcccccccc}
\toprule $TeVCat Name$ & $R.A./^\circ$   & $Dec./^\circ$ & $\sigma$ & $N_0/(TeV^{-1 }cm^{-2}s^{-1})$ &  $E_{0}/TeV$ & $\beta$ & $Extension/^\circ$ & Ref. \\
\hline  
2HWC J1309-054 & 197.31 & -5.49 & 7.8& $1.23\times 10^{-14}$ & 7 & 2.55 & \_\_ & \cite{Abeysekara2017} \\
HESS J1813-126 & 273.34 & -12.69 & 5.9 & $2.74\times 10^{-14}$ & 7 & 2.84 & \_\_ & \cite{Abeysekara2017} \\
2HWC J1825-134 & 276.46 & -13.40 & 8.0 & $2.49\times 10^{-13}$ & 7 & 2.56 & 0.9 & \cite{Abeysekara2017} \\
2HWC J1829+070 & 277.34 & 7.03 &11.1 & $8.10\times 10^{-15}$ & 7 & 2.69 & \_\_ & \cite{Abeysekara2017} \\
2HWC J1837-065 & 279.36 & -6.58 & 35.1 & $3.41\times 10^{-13}$ & 7 & 2.66 & 2 & \cite{Abeysekara2017} \\
2HWC J1844-032 & 281.07 & -3.25 & 10.8 & $9.28\times 10^{-14}$ & 7 & 2.51 & 0.6 & \cite{Abeysekara2017} \\
2HWC J1852+013 & 283.01 & 1.38 & 27.8 & $1.82\times 10^{-14}$ & 7 & 2.9 & \_\_ & \cite{Abeysekara2017} \\
MAGIC J1857.6+0297 & 284.40 & 2.97 & 9.2 & $6.10\times 10^{-12}$ & 1 & 2.39 & 0.1 & \cite{Aharonian2008} \\
HESS J1858+020 & 284.58 & 2.09 & 8.3 & $6.00\times 10^{-13}$ & 1 & 2.17 & 0.08 & \cite{Aharonian2008} \\
2HWC J1902+048 & 285.51 & 4.86 & 31.1& $8.30\times 10^{-15}$ & 7 & 3.22 & \_\_ & \cite{Abeysekara2017} \\
2HWC J1907+084 & 286.79 & 8.50 & 31.6& $7.30\times 10^{-15}$ & 7 & 3.25 & \_\_ & \cite{Abeysekara2017} \\
MGRO J1908+06 & 286.98 & 6.27 &10.9 & $8.51\times 10^{-14}$ & 7 & 2.33 & 0.8 & \cite{Abeysekara2017} \\
2HWC J1914+117 & 288.68 & 11.72 & 20.5 & $8.50\times 10^{-15}$ & 7 & 2.83 & \_\_ & \cite{Abeysekara2017} \\
2HWC J1921+131 & 290.30 & 13.13 &20.9& $7.90\times 10^{-15}$ & 7 & 2.75 & \_\_ & \cite{Abeysekara2017} \\
2HWC J1928+177 & 292.15 & 17.78 & 20.1 & $1.07\times 10^{-14}$ & 7 & 2.6 & \_\_ & \cite{Cornwall2006} \\
2HWC J1938+238 & 294.74 & 23.81 &26.3& $7.40\times 10^{-15}$ & 7 & 2.96 & \_\_ & \cite{Abeysekara2017} \\
2HWC J1953+294 & 298.26 & 29.48 & 21.8 & $8.30\times 10^{-15}$ & 7 & 2.78 & \_\_ & \cite{Abeysekara2017} \\ 
2HWC J1955+285 & 298.83 & 28.59 & 7.8 & $5.70\times 10^{-15}$ & 7 & 2.4 & \_\_ & \cite{Abeysekara2017} \\
2HWC J2006+341 & 301.55 & 34.18 & 119.6 & $9.60\times 10^{-15}$ & 7 & 2.64 & \_\_ & \cite{Abeysekara2017} \\
VER J2019+407 & 305.02 & 40.76 &22.9& $1.50\times 10^{-12}$ & 1 & 2.37 & 0.23 & \cite{Aliu2013b} \\
\bottomrule
\end{tabular*}%
\end{center}

\begin{multicols}{2}

\bibliographystyle{unsrt_update}
\bibliography{mybibfile_new3}

\begin{thebibliography}{10}

\bibitem{Morlino2013}
G.~{Morlino}.
\newblock {\em Nucl. Instrum. Methods Phys. Res., Sect. A}, 720:70--73, Aug
  2013.

\bibitem{Saha2015}
L.~{Saha} and P.~{Bhattacharjee}.
\newblock {\em J. High Energy Astrop.}, 5:9--14, March 2015.

\bibitem{DeJager1992}
O.C. {De Jager} and A.K. Harding.
\newblock {\em ApJ}, 396:161--172, 1992.

\bibitem{Ackermann2012}
M.~Ackermann, M.~Ajello, W.~B. Atwood, et~al.
\newblock {\em ApJ}, 750(1), 2012.

\bibitem{Abramowski2015}
A.~Abramowski, F.~Acero, F.~Aharonian, et~al.
\newblock {\em MNRAS}, 446:1163--1169, 2015.

\bibitem{Aleksic2012}
J.~Aleksic, E.A. Alvarez, L.~A. Antonelli, P.~Antoranz, and M.~Asensio.
\newblock {\em A{\&}A}, 544:A96, 2012.

\bibitem{Allen2017}
C.~{Allen}, S.~{Archambault}, A.~{Archer}, et~al.
\newblock {\em MNRAS}, 471(2):2117--2123, Oct 2017.

\bibitem{Amenomori2005a}
R.~U. {Abbasi}, M.~{Abe}, T.~{Abu-Zayyad}, et~al.
\newblock {\em ApJ}, 804(2):133, May 2015.

\bibitem{Bartoli2014}
B.~Bartoli, P.~Bernardini, X.~J. Bi, et~al.
\newblock {\em ApJ}, 790:152, 2014.

\bibitem{Becker2007}
J.~K. Becker, W.~Bednarek, K.~Berger, et~al.
\newblock {\em ApJ}, 664:L91--L94, 2007.

\bibitem{Abeysekara2017}
A.~U. Abeysekara, A.~Albert, R.~Alfaro, et~al.
\newblock {\em ApJ}, 841:100, 2017.

\bibitem{2009icrc}
Z.~G. {Yao}, M.~{Zha}, Z.~{Cao}, and H.~H. {He}.
\newblock {LHAASO Simulation: Performance of the Water Cherenkov Detector
  Array}.
\newblock In {\em 31th International Cosmic Ray Conference (ICRC2009)},
  International Cosmic Ray Conference, Jul 2009.

\bibitem{Yao}
Z.~G. Yao, H.~R. Wu, M.~J. Chen, B.~Gao, and B.~Zhou.
\newblock {\em ICRC 2011}, 9:95--98, 2011.

\bibitem{Bai2019}
X.~{Bai}, B.~Y. {Bi}, X.~J. {Bi}, et~al.
\newblock {\em arXiv e-prints}, page arXiv:1905.02773, May 2019.

\bibitem{1998cmcc.book.....H}
J.~N. {Capdevielee}, P.~{Gabriel}, H.~J. {Gils}, et~al.
\newblock In {\em Very High Energy Cosmic-Ray Interactions}, volume 276, pages
  545--553, Jun 1993.

\bibitem{2003NuPhS.125...60G}
{Geant4 Collaboration} and M.~G. {Pia}.
\newblock {\em Nucl. Phys. B, Proc. Suppl.}, 125:60--68, Sep 2003.

\bibitem{Bartoli2015}
B.~Bartoli, P.~Bernardini, X.~J. Bi, et~al.
\newblock {\em ApJ}, 798:119, 2015.

\bibitem{2003APh....19..193H}
J{\"o}rg~R. {H{\"o}randel}.
\newblock 19(2):193--220, May 2003.

\bibitem{Hampel-Arias2015}
Z.~{Hampel-Arias} and S.~{Westerhoff}.
\newblock In {\em ICRC2015}, volume~34, page 1001, Jul 2015.

\bibitem{Atkins2003}
R.~Atkins, W.~Benbow, D.~Berley, et~al.
\newblock {\em ApJ}, 595(2):803--811, 2003.

\bibitem{Gaisser2013}
T.~K. Gaisser, T.~Stanev, and S.~Tilav.
\newblock {\em Front. Phys.}, 8(6):748--758, 2013.

\bibitem{Guo2016}
Y.~Q. Guo and Q.~Yuan.
\newblock {\em Phys. Rev. D}, 97(6):063008, Mar 2018.

\bibitem{Aharonian2008}
F.~Aharonian, A.~G. Akhperjanian, U.~Barres~De Almeida, B.~Behera, and
  M.~Beilicke.
\newblock {\em A{\&}A}, 477:353--363, 2008.

\bibitem{Amenomori2005}
M.~Amenomori, S.~Ayabe, D.~Chen, et~al.
\newblock {\em ApJ}, 633:1005--1012, 2005.

\bibitem{1975CoPhC..10..343J}
F.~{James} and M.~{Roos}.
\newblock {Minuit - a system for function minimization and analysis of the
  parameter errors and correlations}.
\newblock {\em Computer Physics Communications}, 10(6):343--367, Dec 1975.

\bibitem{Ackermann_2012}
M.~Ackermann, M.~Ajello, W.~B. Atwood, et~al.
\newblock {\em ApJ}, 750(1):3, apr 2012.

\bibitem{Evoli_2008}
Carmelo Evoli, Daniele Gaggero, Dario Grasso, and Luca Maccione.
\newblock Cosmic ray nuclei, antiprotons and gamma rays in the galaxy: a new
  diffusion model.
\newblock {\em J. Cosmol.Astropart. Phys.}, 2008(10):018, oct 2008.

\bibitem{2018ChPhC..42g5103G}
Y.~Q. {Guo} and Q.~{Yuan}.
\newblock {\em Chin. Phys. C}, 42(7):075103, Jun 2018.

\bibitem{Zhao2016}
Y.~Zhao, Q.~Yuan, X.~J. Bi, F.~R. Zhu, and H.~Y. Jia.
\newblock {\em Int. J. Mod. Phys. D}, 25(1):1650006, Oct 2016.

\bibitem{Archambault2016}
S.~Archambault, A.~Archer, T.~Aune, et~al.
\newblock {\em ApJL}, 817:L7, 2016.

\bibitem{Kumar2015}
S.~{Kumar} and {VERITAS Collaboration}.
\newblock In {\em ICRC2015}, volume~34, page 760, Jul 2015.

\bibitem{Ahnen2016}
M.~L. Ahnen, S.~Ansoldi, L.~A. Antonelli, et~al.
\newblock {\em A{\&}A}, 595:A98, 2016.

\bibitem{Archambault2013}
S~Archambault, T~Arlen, T~Aune, et~al.
\newblock {\em ApJ}, 776(69):(10pp), 2013.

\bibitem{Acciari2010}
V.~A. Acciari, E.~Aliu, T.~Arlen, et~al.
\newblock {\em ApJL}, 715:L49--L55, 2010.

\bibitem{Aliu2013}
E.~Aliu, S.~Archambault, T.~Arlen, et~al.
\newblock {\em ApJ}, 779:92, 2013.

\bibitem{Ollaboration2013}
Arun~S. {Madhavan}.
\newblock page arXiv:1307.7051, Jul 2013.

\bibitem{Aliu2008}
V.A. Acciari, E.~Aliu, M.~Beilicke, W.~Benbow, and S.~M. Bradbury.
\newblock {\em ApJ}, 684:L73--L77, 2008.

\bibitem{Petry2002}
D.~Petry, I.H. Bound, S.~M. Bradbury, et~al.
\newblock {\em ApJ}, 580:104--109, 2002.

\bibitem{Aliu2013yy}
E.~Aliu, S.~Archambault, T.~Arlen, et~al.
\newblock {\em ApJ}, 775(1), 2013.

\bibitem{Benbow2017}
C.~{Allen}, S.~{Archambault}, A.~{Archer}, et~al.
\newblock {\em MNRAS}, 471(2):2117--2123, Oct 2017.

\bibitem{Cornwall2006}
D.~Cornwall and A.~Mattingly.
\newblock {\em ApJ}, 643:L53--L56, 2006.

\bibitem{Aliu2013b}
E.~Aliu, S.~Archambault, T.~Arlen, et~al.
\newblock {\em ApJ}, 770:93, 2013.

\end{thebibliography}

\end{multicols}

\clearpage

\end{CJK*}
\end{document}